\documentclass[runningheads]{llncs}
\usepackage[T1]{fontenc}
\usepackage{graphicx}
\usepackage{amsmath,amssymb}
\usepackage{algorithmic}
\usepackage{textcomp}
\usepackage{xcolor}
\usepackage{calc}
\usepackage{tcolorbox}
\usepackage{multirow}
\usepackage{url}
\usepackage{hyperref}
\usepackage{pgfplots}
\pgfplotsset{compat=1.18}
\usepackage{placeins}
\usepackage{fontawesome}
\usepackage{comment}
\usepackage{tabularx}
\usepackage{enumitem}
\usepackage{booktabs}
\usepackage{framed}
\usepackage{fancybox}
\usepackage{makecell}
\usepackage{csquotes}
\usepackage{orcidlink}
\usepackage{multicol}
\usepackage{colortbl}
 
\usepackage{rotating}

\begin{document}

\title{How Does Users' App Knowledge Influence the Preferred Level of Detail and Format of Software Explanations?}
%
\titlerunning{Influence of User App Knowledge on Software Explanations}
%
\author{
    Martin Obaidi\orcidlink{0000-0001-9217-3934}\inst{1} \and
    Jannik Fischbach\orcidlink{0000-0002-4361-6118}\inst{2,3} \and
    Marc Herrmann\orcidlink{0000-0002-3951-3300}\inst{1} \and
    Hannah Deters\orcidlink{0000-0001-9077-7486}\inst{1} \and
    Jakob Droste\orcidlink{0000-0001-8746-6329}\inst{1} \and
    Jil Klünder\orcidlink{0000-0001-7674-2930}\inst{1} \and
    Kurt Schneider\orcidlink{0000-0002-7456-8323}\inst{1}
}

\authorrunning{Obaidi et al.}

\institute{
    Leibniz University Hannover, Software Engineering Group, Hannover, Germany \\
    \email{\{martin.obaidi, marc.herrmann, hannah.deters, jakob.droste, jil.kluender, kurt.schneider\}@inf.uni-hannover.de} \and
    Netlight Consulting GmbH, Munich, Germany \\
    \email{jannik.fischbach@netlight.com} \and
    fortiss GmbH, Munich, Germany \\
    \email{fischbach@fortiss.org}
}
\maketitle              
\begin{abstract}
Context and Motivation: Due to their increasing complexity, everyday software systems are becoming increasingly opaque for users. A frequently adopted method to address this difficulty is explainability, which aims to make systems more understandable and usable. 
Question/problem: However, explanations can also lead to unnecessary cognitive load. Therefore, adapting explanations to the actual needs of a user is a frequently faced challenge.
Principal ideas/results: This study investigates factors influencing users' preferred the level of detail and the form of an explanation (e.g., short text or video tutorial) in software. We conducted an online survey with 58 participants to explore relationships between demographics, software usage, app-specific knowledge, as well as their preferred explanation form and level of detail. The results indicate that users prefer moderately detailed explanations in short text formats. Correlation analyses revealed no relationship between app-specific knowledge and the preferred level of detail of an explanation, but an influence of demographic aspects (like gender) on app-specific knowledge and its impact on application confidence were observed, pointing to a possible mediated relationship between knowledge and preferences for explanations.
Contribution: Our results show that explanation preferences are weakly influenced by app-specific knowledge but shaped by demographic and psychological factors, supporting the development of adaptive explanation systems tailored to user expertise. These findings support requirements analysis processes by highlighting important factors that should be considered in user-centered methods such as personas.

\end{abstract}

\section{Introduction}
\label{sec:intro}
With the increasing complexity of software systems~\cite{Mens2012complex,Antinyan2020complex,levy2021understanding}, users experience everyday technologies being more opaque and challenging to navigate~\cite{Andrade2019usercomplex,Gabbas2021usercomplex}. A common approach to mitigate this issue is through enhancing explainability, which aims to make systems more understandable and effectively usable. Identifying users' needs for explanations is fundamental to this field~\cite{droste2024explanations,unterbusch2023explanation,kohl2019explainability,Kim2023whatwhenexplain}. However, explanation needs vary individually for each user within a system and can manifest with varying intensity~\cite{chazette2020explainability,droste2023designing,goldstandard-explain-zenodo2024}. 

Building an explainable system requires explanations to be formulated. These explanations range from the unavailability of certain features and the storage of 
personal data to detailed steps for using specific functions, which we all refer to as ``software explanations'', summarizing responses to various user queries. Depending on the user's need, these explanations may take the form of user guides, tutorials, or quick tips. For example, when a user asks \enquote{What does this button do?}, an explanation can be provided to clarify this, e.g., \enquote{This function enables dark mode, which uses more dark colors and fewer bright colors}. However, how detailed and in what form these explanations should be presented to satisfy users has not been extensively investigated. 

Previous studies have shown that unnecessary or lengthy explanations can increase the users' cognitive load, resulting in a negative user experience~\cite{chazette2020explainability,nunes2017systematic}. A brief explanation might be appropriate for an expressed need for clarification, such as \enquote{The function allows you to edit images.} However, users might prefer a detailed answer, such as \enquote{This insert function offers advanced image editing features such as adjusting contrast, brightness, color saturation, and integrating layers. It also allows direct linking to external image sources and applying custom scripts for advanced image manipulation.}

Given the challenge and time-consuming nature of querying many users for their preferred explanations across various explanation needs, it is essential to first examine whether there is a relationship between the preferred level of detail and the form of an explanation and demographic data or other aspects such as the users' app-specific knowledge\footnote{With “app-specific knowledge”, we refer to a user's familiarity with software in different functional areas.}.  
Software companies may have access to users' demographic data and app-specific knowledge within the software (e.g., in the game genre). By examining whether demographic data and app-specific knowledge correlate with the preferred level of detail in explanations, it becomes possible to formulate appropriately adapted explanations if the results indicate such correlations.

To investigate such correlations, we conducted an online survey with 58 participants. We asked them demographic questions, assessed their confidence in using software, and had them self-assess their app-specific knowledge in two exemplary types of software: office software (e.g., Microsoft Office) and browser software (e.g., Mozilla Firefox).
We used correlation analysis to examine the relationships between participants' objective app-specific knowledge, demographic data, and usage confidence with their preferred explanation level of detail and form.
Our results indicate that users generally prefer moderately detailed explanations in short text formats, with only weak correlations observed between app-specific knowledge and explanation preferences. Demographic factors, such as age and gender, showed limited influence, suggesting that explanation needs are largely subjective.

\section{Background and Related Work}
\label{sec:background}

This section presents background information and related work on explainability in software engineering.

In requirements engineering, explainability is viewed as a non-functional requirement~\cite{chazette2020explainability,kohl2019explainability} and a software quality aspect~\cite{chazette2021exploring} that involves software behavior and explanatory elements~\cite{kohl2019explainability}. The need for explanations arises when users or developers face unclear system aspects, requiring careful implementation to avoid cognitive overload~\cite{droste2024explanations,chazette2020explainability}. Additionally, explainability supports ethical standards by clarifying input, training, and regulatory compliance~\cite{buiten2023vision}.
In XAI, explainability (often synonymous with interpretability) aims to clarify AI model reasoning and outputs, enhancing system transparency and trust~\cite{droste2024peeking,kastner2021relation}. Beyond XAI, explainability addresses software interactions, privacy concerns~\cite{brunotte2023privacy}, and app-specific challenges like unfamiliar terminology~\cite{droste2024explanations,deters2023ondemand}.
Obaidi et al. explored whether a user's mood influences the type and frequency of explanation needs~\cite{obaidi2025mood}, finding that explanation preferences are highly subjective and exhibit only weak or no correlations with mood-related factors. 
Deters et al.~\cite{deters2024qualitymodel} developed a quality model that assesses the degree of explainability in software systems.
Unterbusch et al.~\cite{unterbusch2023explanation} emphasize the importance of understanding and classifying users' explanation needs for software systems. They manually analyzed 1,730 app reviews from eight apps to develop a taxonomy of explanation needs and explored automated approaches to detect them. Their best classifier achieved a weighted F-score of 86\% in identifying explanation needs in 486 unseen reviews across four apps.  
Sadeghi et al.~\cite{sadeghi2024explanation} investigate how misleading explanations affect trust in systems performing unreliable and hard-to-assess tasks. Their online survey with 162 participants found that those exposed to misleading explanations rated their trust significantly higher and aligned their predictions more closely with the system's output than those who saw only the system’s predictions. The authors recommend calibrating explanations according to the system’s confidence to mitigate misleading effects.
Deters et al.~\cite{deters2023expl} explore how to evaluate explainability in software systems. 
They discovered that users' prior knowledge should also be included in the evaluation, as users with different prior knowledge require different explanations. However, they did not explore how different levels of prior knowledge could be considered in the generation of explanations.
Ramos et al.~\cite{ramos2021modeling} analyzed how the need for explanation differs for different users by creating so-called personas. These personas reflect user groups and are intended to help requirements analysts identify the need for explanation that these different user groups have for recommender systems. They mainly focus on trust in recommendations and interest in explainability. However, Ramos et al.~\cite{ramos2021modeling} do not yet consider users' familiarity with apps. 
Obaidi et al.~\cite{obaidi2025automatingexplanationneedmanagement} examined the need for explanations in app reviews of a company that develops multiple navigation apps. They developed an automated approach to identify explanation needs within these navigation apps and determine which team within the company is most likely to provide the required information or where the relevant details can be found.

\section{Study Design}
\label{sec:research}
To gain an overview of the app-specific knowledge the users possess and the preferred explanation level of detail they desire, we developed and conducted a survey. 

\subsection{Research Questions}

We investigate the following research questions (RQ):

\noindent \textbf{RQ1: How does app-specific knowledge relate to preferred explanation level of detail and form?} Having objective criteria presents a first step towards adapting explanations to the users' needs. 

\noindent \textbf{RQ2: How does confidence in using software relate to preferred explanation level of detail and form?} Answering this question can help tailor support materials to user needs, enhancing user satisfaction and efficiency. Software developers can reduce cognitive load by providing explanations that match users' confidence levels.

\noindent \textbf{RQ3: How does demographic data relate to preferred explanation level of detail and form?} Understanding how demographic data relates to explanation needs enables the customization of support materials, ensuring inclusivity and accessibility. Tailoring explanations based on demographic factors such as age, education, and professional background enhances user satisfaction and effective software use.

\subsection{Survey Structure}
Table~\ref{tab:survey-structure} shows an overview of our survey questions. To avoid the mono-operation bias, we considered two types of software, namely office software (such as Microsoft Office) and browser software (such as Mozilla Firefox), as we assume a wide variety in the user groups of these apps that can be helpful in the context of our study. Unlike many AI-focused explainability studies, our work examines everyday software, where users primarily need explanations for interaction rather than AI functions~\cite{droste2024explanations,goldstandard-explain-zenodo2024}. By focusing on widely used applications, we aim to provide insights for improving practical user support and interface design. Our study follows the concurrent control design without random assignment as described by Kitchenham and Pfleeger~\cite{kitchenham2002survey2}.

\begin{table*}[htbp]
\footnotesize
    \caption{Survey structure overview. [App category] represents Browser or Office.}
    \centering
    \begin{tabularx}{\textwidth}{Xl}
    \toprule
    \textbf{Questions and Answer Options} & \textbf{Variable} \\ 
    \midrule
    \textit{Demographics} & \\ \midrule
    Gender? (e.g., female, male, diverse) & $d\textsubscript{gender}$ \\
    Age? (18-99) & $d\textsubscript{age}$ \\
    Field of qualifications? (e.g., Education, Finance, IT, etc.) & $d\textsubscript{job}$ \\
    \midrule 
    \textit{{[App category]} Software Usage} & \\ \midrule
    Software used for work? (e.g., Chrome, Word) & - \\
    Rate your knowledge in {[app category]} software? (Very low to Very high) & $C(f_{1}, d)$ \\
    Objective knowledge test (6 questions) & $C(f_{2}, d)$ \\
    \midrule
    \textit{{[App category]} Software Experience} & \\ \midrule
    How often do you feel confident in {[app category]} software?  & $S\textsubscript{sw}$ \\
    Preferred explanation form? (e.g., Text, Video, Training, etc.) & $E\textsubscript{form}$ \\
    Desired explanation detail level? (Very little to Extremely detailed) & $E\textsubscript{detail}$ \\
    \midrule
    \textit{Opinions on {[App category]} Software} & \\ \midrule
    Satisfaction with user-friendliness? (Very dissatisfied to Very satisfied) & - \\
    \bottomrule
    \end{tabularx}
    \label{tab:survey-structure}
\end{table*}

\subsubsection{Explanation Level of Detail and Form}
\label{sec:detailliertheit}

This work focuses on the concept of preferred explanation level of detail and form and their relationships with app-specific knowledge and other variables. Explanation texts, fundamental to software, are central to this study. They convey an understanding of individual functions and can describe anything from a button to an entire program flow. Explanation texts vary in their level of detail and are here assumed as textual explanations. The levels of detail are defined not only by text length, but also by the type and amount of information provided, ranging from superficial to highly detailed information. In this study, we define five levels of detail, each illustrated with examples presented in the survey:

\begin{enumerate}
    \small
    \item \textbf{Very low detail:} Texts provide only basic information and are compact. \textit{Example:} The function allows inserting images into a document.
    \item \textbf{Low detail:} Texts offer additional tips but may still leave gaps in understanding. \textit{Example:} This function allows you to insert images into your document to make it visually more appealing.
    \item \textbf{Detailed:} Texts provide comprehensive descriptions, including examples. \textit{Example:} The insert function enables adding images of various formats and sizes to the document. You can adjust the position, size, and alignment of the images.
    \item \textbf{Highly detailed:} Texts delve deeply into software functionality, sharing technical details. \textit{Example:} Within this function, you can make detailed adjustments to inserted images, including image captions, text wrapping, alignment options, and filter effects.
    \item \textbf{Very highly detailed:} Texts offer extensive explanations, covering all technical aspects, interfaces, extensions, and terminology. \textit{Example:} This insert function offers advanced image editing features such as adjusting contrast, brightness, color saturation, and integrating layers. It also allows direct linking to external image sources and applying custom scripts for advanced image manipulation.
\end{enumerate}

The form of explanation, like the level of detail, is divided into five levels: 
\begin{enumerate}
    \small
    \item No specific form or others.
    \item Simple form: e.g., short text instructions.
    \item More complex form: e.g., video tutorials (GIFs).
    \item Elaborate form: e.g., interactive training.
    \item Very time-consuming form: e.g., live chat support.
\end{enumerate}
The form of explanation increases in detail and complexity, similar to the level of detail. Note that during the online survey, each explanation detail level was accompanied by an example to ensure clarity for participants.

\subsubsection{App-specific Knowledge}
\label{sec:domeanenwissen}

We aimed to capture how well participants understood software-specific terminology and functionalities within these app categories. The study scales app-specific knowledge into five levels:
\begin{enumerate}
    \small
    \item Very low: Minimal familiarity and experience with the software.
    \item Low: Limited experience with a few successes.
    \item Average: Most users fall into this category, able to use the software effectively but requiring external help for deeper functions.
    \item High: Extensive experience, able to complete complex tasks independently.
    \item Very high: Expert-level understanding, rarely encountering problems and capable of assisting average users.
\end{enumerate}

\subsection{Data Collection}

The survey was hosted using the institutional \href{https://www.limesurvey.org/}{LimeSurvey} server. 
We distributed the survey via email and social media apps like WhatsApp or Signal to personal contacts and groups. While this distribution method might limit external validity, it ensured a range of participant backgrounds. We invited potential software users who were of legal age (over 18). The survey was conducted in February 2024. In total, 83 participants took part, with 58 completing all survey sections, while we have 57 responses for the browser app category and 53 for the office app category~\cite{obaidi_2024_correlation-apps}.

\subsection{Data Analysis}
\label{sec:analyseaufbau}

To use the survey data properly in the correlation analysis, specific excerpts were extracted.
An overview of the variables used in our hypotheses and their mapping to the survey questions can be found in Table~\ref{tab:vars-uebersicht}. 

\begin{table*}[htb]
    \centering
    \footnotesize
    \setlength{\tabcolsep}{3pt} 
    \caption{Variables Overview. Abbreviation: expl.=explanation, Prof.=Professional.}
    \label{tab:vars-uebersicht}
    \begin{tabularx}{\textwidth}{lXllX}
        \toprule
        \textbf{Variable}   & \textbf{Survey Question} & \textbf{Scale} & \textbf{Range} & \textbf{Description} \\ 
        \midrule
        $C\textsubscript{sub}$ & Subj. app knowledge & Ordinal & 1..5 & Self-assessed knowledge \\ 
        $C\textsubscript{obj}$ & Obj. app knowledge & Ordinal & 1..5 & Assessed through quiz \\ 
        $S\textsubscript{sw}$ & Confidence using apps & Ordinal & 1..5 & \\ 
        $E\textsubscript{form}$ & Preferred expl. form & Nominal &  & Expl. representation \\ 
        $E\textsubscript{detail}$ & Explanation detail levels & Ordinal & 1..5 & Detail level of expl. text   \\ 
        $d\textsubscript{gender}$ & Gender? & Nominal & \{1; 2; 3\} & Female, male, diverse \\ 
        $d\textsubscript{age}$ & Age & Metric & $n \in \mathbb{N\textsubscript{0}}$ & \\ 
        $d\textsubscript{job}$ & Prof. qualification field & Nominal & \{1; ... ; 8\} & \\ 
        \bottomrule
    \end{tabularx}
\end{table*}

\subsubsection{Determining Objective App-specific Knowledge}

To objectively assess app-specific knowledge in the office and browser app categories, participants were asked to answer a series of questions. Each app category had six questions with four possible answers each, with participants also having the option to indicate if they did not know the answer. The questions included multiple correct answers. In total, there were ten correct answers for the office app category and eight for the browser app category. We calculated the number of correct answers, and based on the percentage of correct responses, we determined their app-specific knowledge according to the categories (Very low <20\% - Low - 40\% - Average - 60\% - High - 80\% - Very high - 100\%).

This way, we were able to compare objectively determined app-specific knowledge with the self-reported subjective app-specific knowledge.

Here are three of the six questions designed to assess objective app-specific knowledge in the area of browsers:
\begin{itemize}
    \small
    \item \enquote{What is the difference between a search engine and a web browser?}
    \item \enquote{What is true about the developer tools in most browsers?}
\end{itemize}

Similarly, here are two questions in the area of office applications.

\begin{itemize}
    \small
    \item \enquote{What basic formatting options are available in most word processing programs?}
    \item \enquote{What advantage do macros provide in Microsoft Excel?}
\end{itemize}

\subsubsection{Hypotheses Testing}
\label{subsec:hypothesen}

To further examine relationships between app-specific knowledge, demographics, and explanation detail level and form, we formulated the hypotheses summarized in Table~\ref{tab:hypothesen-uebersicht}. 

\begin{table*}[thb]
\centering
\small
    \setlength{\tabcolsep}{4pt} 
\caption{Overview of Hypotheses. AK=app-specific knowledge, "/"=or.}
\label{tab:hypothesen-uebersicht}
\begin{tabularx}{\textwidth}{lXl}
\toprule
\multicolumn{3}{c}{\textbf{Hypotheses}} \\ \midrule
H1\textsubscript{0}   & \textbf{Self-assessed AK is unrelated to objective AK.}                                    &         \\
H1.1\textsubscript{0} & Self-assessed AK is unrelated to objective AK in Office.                &         \\
H1.2\textsubscript{0} & Self-assessed AK is unrelated to objective AK in Browser.               &         \\ \hline
H2\textsubscript{0}   & \textbf{Obj. AK does not impact explanation form / detail level.} & RQ1       \\
H2.1\textsubscript{0} & Objective AK does not impact explanation form in Office.              & RQ1       \\
H2.2\textsubscript{0} & Objective AK does not impact explanation form in Browser.             & RQ1       \\
H2.3\textsubscript{0} & Objective AK does not impact explanation detail level in Office.              & RQ1       \\
H2.4\textsubscript{0} & Objective AK does not impact explanation detail level in Browser.             & RQ1       \\ \hline
H3\textsubscript{0}   & \textbf{Confidence does not impact explanation form / detail level.} & RQ2       \\
H3.1\textsubscript{0} & Confidence does not impact form in Office.              & RQ2       \\ 
H3.2\textsubscript{0} & Confidence does not impact form in Browser.              & RQ2       \\
H3.3\textsubscript{0} & Confidence does not impact detail level in Office.                 & RQ2       \\
H3.4\textsubscript{0} & Confidence does not impact detail level in Browser.                & RQ2       \\ \hline
H4\textsubscript{0}   & \textbf{Demographics do not impact explanation form / detail level.}                           & RQ3         \\
H4.1\textsubscript{0} & Gender does not impact form in Office.                    & RQ3         \\
H4.2\textsubscript{0} & Gender does not impact form in Browser.                    & RQ3         \\
H4.3\textsubscript{0} & Gender does not impact detail level in Office.                    & RQ3         \\
H4.4\textsubscript{0} & Gender does not impact detail level in Browser.                   & RQ3         \\
H4.5\textsubscript{0} & Age does not impact form in Office.                    & RQ3         \\
H4.6\textsubscript{0} & Age does not impact form in Browser.                    & RQ3         \\
H4.7\textsubscript{0} & Age does not impact detail level in Office.                    & RQ3         \\
H4.8\textsubscript{0} & Age does not impact detail level in Browser.                   & RQ3         \\
H4.9\textsubscript{0} & Prof. experience does not impact form in Office.                    & RQ3         \\
H4.10\textsubscript{0} & Prof. experience does not impact form in Browser.                    & RQ3         \\
H4.11\textsubscript{0} & Prof. experience does not impact detail level in Office.                    & RQ3         \\
H4.12\textsubscript{0} & Prof. experience does not impact detail level in Browser.                   & RQ3         \\ \midrule
\multicolumn{3}{c}{\textbf{Variable Relationships}}                                                          \\ \midrule
H1   & \multicolumn{2}{l}{App-specific knowledge (\textit{AK})}                                                                             \\
H2   & \multicolumn{2}{l}{\makecell[l]{Objective AK (\textit{C($f_{2}$, d)}), explanation form (\textit{E\textsubscript{form}}), detail level (\textit{E\textsubscript{detail}}),\\app categories (\textit{d})}}                                                                 \\
H3   & \multicolumn{2}{l}{Confidence (\textit{S\textsubscript{sw}}), form (\textit{E\textsubscript{form}}), detail level (\textit{E\textsubscript{detail}}), app categories (\textit{d})}                                                                 \\
H4   & \multicolumn{2}{l}{Gender (\textit{d\textsubscript{gender}}), age (\textit{d\textsubscript{age}}), prof. experience (\textit{d\textsubscript{job}}),}                                                                                                                    \\
     & \multicolumn{2}{l}{explanation form (\textit{E\textsubscript{form}}), detail level (\textit{E\textsubscript{detail}}), app categories (\textit{d})}              \\ 
     \bottomrule
\end{tabularx}
\end{table*}

We apply the Bonferroni correction at the hypothesis level rather than the sub-hypothesis level to manage the potential for Type I errors effectively~\cite{haynes2013bonferroni}. For instance, with four subordinate hypotheses, the corrected significance level is $\alpha\textsubscript{corr} = \frac{0.05}{4} = 0.0125$. 

In this study, correlations for nominal (categorical) data are examined using the Pearson $\chi^2$ test, while the Spearman correlation is employed for ordinal or metric data. For the Pearson $\chi^2$ test, the calculated statistic is interpreted via Cramer's V and the contingency coefficient (CC) to assess the strength of the association. Conversely, for the Spearman correlation, Cohen's effect size is used to evaluate the magnitude of the relationship. Thus, the $\chi^2$ result (via Cramer's V/CC) and the correlation coefficient $r\textsubscript{s}$ serve as the basis for evaluating the effect size of the respective relationships.

\section{Results}
\label{sec:results}
All study data and our code are publicly available at Zenodo \cite{obaidi_2024_correlation-apps}.

\subsection{Population and Survey Sample}
\label{sec:randdaten}

58 participants took part in the survey, including 11 women, 46 men, and one person identifying as non-binary. The average age is 34.1 years, with the youngest participant being 18 years old and the oldest 84 years old. The median is 25.5 years and the standard deviation is 17.8. The participants work in the following industries: IT/technology (13), education (9), finance (3), legal fields (3), healthcare (1) and other (8). Additionally, participants could indicate if they were students (28). If no professional field was applicable or if a participant was not a student, the option "none" was available (1). Participants could also specify other professional fields (8), including media designer for image and sound, electrical engineering, and architecture. 

Additionally, it was determined that five students already have a degree or are working in a professional field. In both app categories, software products were provided as orientation aids. Participants were asked to check which products they use in their daily lives or at work. If "none" was checked, the survey ended for those participants, and no further data was collected. This did not happen in any of the 58 data points used in this study.

Participants reported the software products they use in both the Browser and Office app categories. In the Browser app category (n=57), the most commonly used software was Firefox (36 users), followed by Chrome (27), Safari (23), Edge (12), Opera (10), and Internet Explorer (4). Six participants reported using other browsers, and one indicated using none.
In the Office app category (n=53), the Microsoft Office Suite (Word, Excel, PowerPoint, etc.) was the most frequently used (40 users), with 27 participants using industry-specific software, 25 using Google Workspace (Docs, Sheets, Slides), and 15 using project management tools (such as Jira or Trello). Six participants reported using other types of software, and five indicated they do not use any Office software.

\subsection{Survey Answers}
Of the total 58 participants, five did not provide any information for the Office app category, resulting in \(n=53\) and analogous \(n=57\) for the Browser app category. In terms of app-specific knowledge, both subjective and objective measures for the Office app category yielded a median rating of 4 (on a scale from 1 to 5), with scores ranging from 1 to 5. For the Browser app category, subjective knowledge also held a median of 4, but objective knowledge reached a median of 5, with minimum and maximum values spanning from 1 to 5 in both subjective and objective assessments.

The preferred explanation detail level and the preferred form for an explanation can be seen in Table~\ref{tab:merged_levels_forms}. 

\begin{table*}[htbp]
    \centering
    \small
    \caption{Frequency of Feedback on Detail Levels and Preferred Forms of Explanations in Office (n=53) and Browsers (n=57)}
    \label{tab:merged_levels_forms}
    \begin{tabularx}{\textwidth}{Xrr@{\hskip 10pt}Xrr}
        \toprule
        \textbf{Detail Level} & \textbf{Browser} & \textbf{Office} & \textbf{Form of Explanation} & \textbf{Browser} & \textbf{Office} \\ 
        \cmidrule(rr){1-3} \cmidrule{4-6}
        Very little detailed     & 11 & 9  & Short text instructions  & 45 & 38 \\
        Little detailed          & 14 & 8  & Video tutorials (GIFs)   & 25 & 32 \\
        Detailed                 & 23 & 29 & Interactive training     & 1  & 0  \\
        Very detailed            & 7  & 5  & Live chat support        & 1  & 4  \\
        Extremely detailed       & 2  & 2  & None                     & 7  & 4  \\
                                 &    &    & Other                    & 2  & 1  \\ 
        \bottomrule
    \end{tabularx}
\end{table*}

The majority of participants prefer moderately detailed explanations for both app categories. Notably, there is a stronger tendency towards less detailed explanations in the Browser app category. This preference is also reflected in the preferred explanation format, where the inclination towards short text explanations is significantly higher for the Browser app category compared to the Office app category.

There were a total of three participants who selected both "none" and another explanation form. Presumably, there are scenarios for these participants where they do not want an explanation as well as scenarios where they want explanations. Alternatively, they may have misunderstood the question or made a mistake. One of these participants also selected "other" for the Browser app category and expressed a desire for there to be no "change page" when an update occurs.

\subsection{Correlation Analysis}
\label{sec:ergebnisse-der-korrelationsanalyse}

The statistical results of the correlation analysis are derived from the tested null hypotheses and can be seen in Table~\ref{tab:auswertung-korrelationsanalyse}.  

We always tested the two variables, \textit{Var\textsubscript{1}} and \textit{Var\textsubscript{2}} for a relationship. 
The hypotheses (see Table~\ref{tab:hypothesen-uebersicht}) are tested for both the office and browser app categories. The variables are taken from the variable overview (see Table~\ref{tab:vars-uebersicht}).

\begin{table*}[ht]
    \centering
    \small
    \setlength{\tabcolsep}{4pt} 
    \caption{Results of the Correlation Analysis. Abbreviations: \textbf{App}=App category, \textbf{Eval.}=Evaluation, \textbf{Relat.}=Relationship, \textbf{p}=p-value, \textbf{not reject}=do not reject.}
    \label{tab:auswertung-korrelationsanalyse}
    \begin{tabularx}{\textwidth}{XXXXrrXX}
        \toprule
        \textbf{H\textsubscript{0}} & \textbf{App} & \textbf{Var\textsubscript{1}} & \textbf{Var\textsubscript{2}} & \textbf{$\boldsymbol{|r\textsubscript{s}|} \backslash \boldsymbol{X^2}$} & \textbf{p} & \textbf{Eval.} & \textbf{Relat.} \\ \midrule
        H1.1\textsubscript{0} & Office & C\textsubscript{sub} & C\textsubscript{obj} & 0.62 & 0.00 & reject & strong\\
        H1.2\textsubscript{0} & Browser & C\textsubscript{sub} & C\textsubscript{obj} & 0.57 & 0.00 & reject & strong\\ \hline
        H2.1\textsubscript{0} & Office & C\textsubscript{obj} & E\textsubscript{form} & -0.49 & 0.00 & reject & moderate\\
        H2.2\textsubscript{0} & Browser & C\textsubscript{obj} & E\textsubscript{form} & -0.11 & 0.41 & not reject & weak\\
        H2.3\textsubscript{0} & Office & C\textsubscript{obj} & E\textsubscript{detail} & -0.07 & 0.61 & not reject & no effect\\
        H2.4\textsubscript{0} & Browser & C\textsubscript{obj} & E\textsubscript{detail} & -0.03 & 0.85 & not reject & no effect\\ \hline
        H3.1\textsubscript{0} & Office & S\textsubscript{sw} & E\textsubscript{form} & -0.34 & 0.01 & reject & moderate\\
        H3.2\textsubscript{0} & Browser & S\textsubscript{sw} & E\textsubscript{form} & -0.10 & 0.47 & not reject & no effect\\
        H3.3\textsubscript{0} & Office & S\textsubscript{sw} & E\textsubscript{detail} & -0.03 & 0.81 & not reject & no effect\\
        H3.4\textsubscript{0} & Browser & S\textsubscript{sw} & E\textsubscript{detail} & 0.12 & 0.38 & not reject & weak\\ \hline
        H4.1\textsubscript{0} & Office & d\textsubscript{gender} & E\textsubscript{form} & 7.96 & 0.44 & not reject & low\\
        H4.2\textsubscript{0} & Browser & d\textsubscript{gender} & E\textsubscript{form} & 12.31 & 0.14 & not reject & moderate\\
        H4.3\textsubscript{0} & Office & d\textsubscript{gender} & E\textsubscript{detail} & 8.92 & 0.35 & not reject & low\\
        H4.4\textsubscript{0} & Browser & d\textsubscript{gender} & E\textsubscript{detail} & 3.28 & 0.92 & not reject & low\\
        H4.5\textsubscript{0} & Office & d\textsubscript{age} & E\textsubscript{form} & -0.06 & 0.66 & not reject & no effect\\
        H4.6\textsubscript{0} & Browser & d\textsubscript{age} & E\textsubscript{form} & -0.08 & 0.54 & not reject & no effect\\
        H4.7\textsubscript{0} & Office & d\textsubscript{age} & E\textsubscript{detail} & 0.14 & 0.30 & not reject & weak\\
        H4.8\textsubscript{0} & Browser & d\textsubscript{age} & E\textsubscript{detail} & 0.04 & 0.78 & not reject & no effect\\
        H4.9\textsubscript{0} & Office & d\textsubscript{job} & E\textsubscript{form} & 16.74 & 0.95 & not reject & low\\
        H4.10\textsubscript{0} & Browser & d\textsubscript{job} & E\textsubscript{form} & 14.67 & 0.98 & not reject & low\\
        H4.11\textsubscript{0} & Office & d\textsubscript{job} & E\textsubscript{detail} & 11.78 & 1.00 & not reject & low\\
        H4.12\textsubscript{0} & Browser & d\textsubscript{job} & E\textsubscript{detail} & 14.67 & 0.98 & not reject & low\\ 
        \bottomrule
    \end{tabularx}
\end{table*}

After applying the Bonferroni correction, Hypotheses H1, H2, and H3 were evaluated for significance, resulting in rejection for each, as at least one of the subordinate hypotheses underscores the adjusted significance level. Specifically, H2.1 and H3.1 were identified as significant among their respective sub-hypotheses. In contrast, H4 was not rejected, as its p-value is 0.0042, thus falling below the required significance level for rejection.

Based on the results of the hypotheses testing, we can conclude: 

\textbf{H1\textsubscript{0}:} The participants' self-assessment closely matches the objectively determined app-specific knowledge. Cohen's effect size also shows a strong relationship. Therefore, H1\textsubscript{0} and all subordinate null hypotheses are rejected.

\textbf{H2\textsubscript{0}:} Objective app-specific knowledge correlates only in the office app category with the preferred explanation form. The p-value of 0.00009 is so low that the correlation is also significant for $\alpha\textsubscript{corr} = 0.0025$. Thus, H2.1\textsubscript{0} is rejected. The other null hypotheses for this research question do not correlate. 

\textbf{H3\textsubscript{0}:} Confidence in using software correlates significantly with the preferred explanation form in the office category for $\alpha = 0.05$ and $\alpha\textsubscript{corr} = 0.0125$. H3.1\textsubscript{0} is rejected. All other subordinate null hypotheses from H3\textsubscript{0} are not rejected.

\textbf{H4\textsubscript{0}:} Demographic data show no statistical significance in any hypothesis. In the browser app category, there is no significance between gender and the preferred explanation form, but a moderate relationship is indicated by Cramer's V. Age has a weak correlation with the preferred explanation detail level in the office app category. Significance for demographic data is not present even without Bonferroni correction. Therefore, the null hypothesis H4\textsubscript{0} is not rejected.

\section{Discussion}
\label{sec:discussion}

In the following, we answer the research questions, present threats to validity, and interpret the results.

\subsection{Answering the Research Questions}
\label{sec:beantworten-der-forschungsfragen}

\textbf{RQ1: }
The correlation analysis indicates no significant correlation between explanation detail level and app-specific knowledge. There is a moderate correlation between the preferred explanation form and app-specific knowledge in the office app category. However, since "app-specific knowledge" pertains to general software familiarity rather than specialized industry expertise, this moderate correlation is not sufficient to establish a generalizable connection between app-specific knowledge and the preferred form or detail level of explanations. Thus, app-specific knowledge, as defined in this study, is only weakly related to the preferred form or detail level of explanations. Nevertheless, our results indicate a relationship that should be examined in further studies. 

\noindent \textbf{RQ2: }
Confidence in using software has a similar relationship with the preferred form or detail level of explanations as app-specific knowledge. Only the preferred explanation form in the office app category shows a moderate correlation. This finding suggests that confidence may play a minor role in determining explanation preferences but does not strongly influence the detail level or form overall.

\noindent \textbf{RQ3: }
There is no significant correlation between demographic data and the preferred form or detail level of explanations. Moreover, if there is a weak to moderate, the preferred form or detail level of explanations is weakly related to factors like age or work experience. These findings suggest that demographic data alone does not serve as strong predictor of users' explanation needs. Future studies could consider psychological factors or specific educational backgrounds as potentially more predictive factors in tailoring explanation preferences.

\subsection{Interpretation}
\label{sec:interpretation}

The highly significant correlation between subjective and objective app-specific knowledge (H1\textsubscript{0}), with a strong effect size, clearly shows that participants have accurately assessed their app-specific knowledge. This correlation supports the reliability of using self-assessed knowledge measures and validates the construct of app-specific knowledge in this study.

The majority of participants prefer short and less detailed explanations, indicating that overly extensive explanations may overwhelm users, which is in line with the assumption from previous work~\cite{chazette2020explainability,nunes2017systematic}. This highlights the importance of designing explanations that are concise yet informative. The study’s findings suggest that users generally favor brief textual explanations over more interactive formats, likely due to time constraints or the simplicity of text for quick comprehension. This preference suggests practical applications in designing support features, where prioritizing brief text instructions could enhance usability and user satisfaction.

The correlation analysis reveals that the preferred form or detail level of explanations does not strongly correlate with any of the tested variables. This result implies that factors beyond those controlled in this study might be more influential in determining explanation preferences. For example, task relevance might impact whether users seek more detailed information, especially in high-stakes software applications (e.g., tax software) where accuracy is critical. Furthermore, individual psychological traits, such as personality types or specific cognitive preferences, could play a substantial role in determining explanation needs. Exploring these and other contextual influences on explanation preferences would be valuable for future research.

An unclear UI that confuses users or causes difficulties in understanding functionalities~\cite{chazette2020explainability} could also impact the preferred explanation detail level or form. Complex algorithms or underlying business logic might prompt users to request deeper explanations to clarify how specific system behaviors relate to their tasks. Additionally, technical issues or unexpected behaviors may lead users to seek more informative explanations to diagnose and resolve errors~\cite{Wiegand2020unexpected}.

App-specific knowledge only weakly to moderately relates to the preference for explanation detail level or form, with a moderate correlation in the office app category. This suggests that users with higher app-specific knowledge may indeed engage in more complex tasks that require concise yet specific explanations, avoiding overly detailed content. Such users likely prefer explanations that directly address advanced functionalities or particular issues rather than generalized overviews. This relationship between confidence in software usage and preferred explanation style underlines the potential of adjusting explanations based on users' app-specific knowledge or software confidence levels, thereby enhancing user satisfaction by aligning explanations with user expertise.

Lastly, textual explanations are consistently preferred across both app categories. Most participants favor reading short explanatory texts over interactive formats. Although training sessions or videos could improve understanding through practical examples, they may conflict with users’ time constraints, making concise texts a more practical choice. This finding supports prioritizing short, straightforward explanation texts as a standard option in user help design, while reserving interactive elements for cases where deeper engagement is necessary.

These findings provide insights for user-centered requirements engineering processes. On the one hand, our work shows that shorter and less detailed explanations are generally preferred, which can be taken into account in elicitation processes. On the other hand, we gain a better understanding of key factors that influence explanation preferences, which can be used for methods such as personas. Characteristics such as app specific knowledge and confidence in the system have a moderate influence on explanation preferences, making them suitable for inclusion in personas.  At the same time, however, the findings also demonstrate that this influence is rather small, leading to the assumption that there might be other important influencing factors. This increases the incentive for further research that examines other factors such as mood or technical background.

\subsection{Future Work}
\label{sec:ausblick}

Future research should aim to include a larger dataset and consider multiple app categories in the online study to enhance generalizability. Expanding the survey to a broader population segment would allow for more representative conclusions, while expert input on app category selection and objective app-specific knowledge assessment could improve measurement validity. Additionally, incorporating a wider range of variables—such as sentiment data or personality traits—would provide a more nuanced understanding of factors influencing explanation preferences.

To further refine these insights, longitudinal studies could explore how explanation preferences evolve as users gain experience with software, enabling the development of dynamic, adaptable explanations. Experimental approaches, such as A/B testing in real-world software environments, could yield empirical evidence on the most effective explanation styles for different user profiles.

Integrating direct interaction with explanation formats within the survey would also enhance ecological validity, allowing participants to experience various explanation types before stating their preferences. This would provide a more accurate reflection of user needs, reducing biases associated with hypothetical scenarios. Additionally, adopting a quasi-experimental approach could offer more controlled conditions, minimizing external biases and enabling a more precise assessment of how app-specific knowledge influences explanation preferences.

Finally, future studies could investigate the practical impact of explanation detail levels and formats on user performance and satisfaction. Measuring task completion times, error rates, and user satisfaction in real-world applications would provide valuable insights into the effectiveness of tailored explanations, further informing user-centered software design.

\subsection{Threats to Validity}
\label{sec:validiteat}

The following section applies the "Threats to Validity" as described by Wohlin~\cite{wohlin2012experimentation} to the content of this work. These threats are categorized into \textit{construct}, \textit{internal}, \textit{conclusion}, and \textit{external} validity.

\textbf{Construct Validity.} The self-established evaluation metric for objective app-specific knowledge and all ordinal scales were only developed by two researchers and were not reviewed by a third instance. Despite multiple reviews of the resulting data, it is possible that errors were overlooked, influencing the results. Additionally, the limitation of variables in the correlation analysis may distort the findings. Although the variables were chosen with a focus on explainability, it remains unclear which factors truly influence the preferred explanation detail level or form. The absence of direct interaction with software explanations in the survey introduces a risk of response bias, as participants may not have had sufficient context to make fully informed responses about their preferred explanation formats. Furthermore, the limited number of app categories restricts the ability to make generalizable statements about explanation preferences across different software types. The selection of these categories could significantly impact the final results.

\textbf{Internal Validity.} Since the study was conducted online, participants could have looked up the answers to the objective knowledge assessment questions, which may have artificially inflated their scores. Additionally, the survey presented examples to clarify different explanation levels and forms, but the lack of direct interaction with software explanations may have still led to variability in participant responses.

\textbf{Conclusion Validity.} The Bonferroni correction was applied to reduce the risk of Type I errors (false positives), but this also increases the risk of Type II errors (false negatives). However, the correction was applied at the hypothesis level to balance these effects, and in cases of observed correlations, it did not alter the overall evaluation.

\textbf{External Validity.} The demographic composition of the survey participants does not represent the general population. The average participant age is 34.1 years, with a gender distribution of 11 women, 46 men, and one non-binary individual, which is not globally representative. Additionally, 48\% of participants are students, limiting the generalizability of findings to professional environments. The sample size of 58 participants further restricts generalization. Furthermore, the study was conducted in only one country, introducing potential cultural biases. These factors suggest the need for broader, more diverse sampling in future research.

\section{Conclusion}
\label{sec:conclusion}

Our study aimed to investigate the relationship between app-specific knowledge, demographics, and the preferred explanation detail level and form. To achieve this, an online survey focused on two app categories was conducted. Participants were asked about the software they frequently use, their self-assessed app-specific knowledge, and the frequency and level of detail of explanations they prefer for the mentioned software. To effectively evaluate the participants' subjective app-specific knowledge, short questions about software in the Office and Browser app categories were posed to gauge their actual understanding in these areas. A correlation analysis was performed to examine the relationships and connections in the survey results. Our analysis revealed no strong relationship between app-specific knowledge and the preferred form or level of detail of explanations. The preferred explanation detail level was categorized into five distinct types, indicating that the preferred explanation detail level either varies between app categories or individuals have a fundamentally consistent preferred explanation level of detail. This suggests a need for flexibility in tailoring explanations, as no "one-size-fits-all" solution was evident. The correlation also revealed that demographic data influence app-specific knowledge and that increased app-specific knowledge correlates with greater application confidence, allowing users to prefer less detailed explanations. Additionally, it was found that participants generally prioritize less detailed explanation texts as their preferred form, likely due to ease of use and time efficiency. 

\section*{Acknowledgment}
This work was funded by the Deutsche Forschungsgemeinschaft (DFG, German Research Foundation) under Grant No.: 470146331, project softXplain (2022-2025).
\subsubsection*{Disclosure of Interests}
{\fontsize{9pt}{11pt}\selectfont The authors have no competing interests to declare that are relevant to the content of this article.}
%
%
%
\bibliographystyle{splncs04}
\bibliography{references.bib}

\end{document}